\documentclass[aps,prl,reprint,superscriptaddress]{revtex4-2}

\usepackage[T1]{fontenc}
\usepackage{graphicx}% Include figure files
\usepackage{dcolumn}% Align table columns on decimal point
\usepackage{bm}% bold math
\usepackage{amsmath,amssymb}
\usepackage{physics} % The physics package provides \ket, \bra, \abs, \dd, etc.
\usepackage[
    colorlinks=true,
    linkcolor=blue,
    citecolor=blue,
    urlcolor=blue
]{hyperref}
\newcommand{\iu}{\mathrm{i}} % Imaginary unit

\begin{document}

%\preprint{APS/123-QED}

\title{Non-Hermitian Theory of Surface Emitting Jackiw-Rebbi States}

\author{Hangkai Fan}
%\email{hangkai.fan@foxmail.com}
\affiliation{Qingdao Innovation and Development Center, Harbin Engineering University, Qingdao 266000, Shandong, China}

\author{Mingzhao Song}
\email{kevinsmz@foxmail.com}
\affiliation{Qingdao Innovation and Development Center, Harbin Engineering University, Qingdao 266000, Shandong, China}

\author{Andrey Bogdanov}
\email{a.bogdanov@metalab.ifmo.ru}
\affiliation{Qingdao Innovation and Development Center, Harbin Engineering University, Qingdao 266000, Shandong, China}
\affiliation{Faculty of Physics, ITMO University, St.\ Petersburg 197101, Russia}

\date{\today}
%\newpage
\begin{abstract}
Topological interface states in open photonic systems combine spatial localization at domain walls with temporal confinement characterized by a radiative quality factor. Here, we develop a non-Hermitian Dirac formalism for surface-emitting Jackiw-Rebbi states in dielectric metagratings. We show how varying the local amplitude of the lattice perturbation shapes the complex Dirac mass profile, enabling control over the quality factors and far-field radiation patterns of the interface states, while preserving their topological character. Applying this framework to a metagrating with second-order Brillouin-zone folding, we design a spatially graded superlattice perturbation that supports interface states with quality factors exceeding 10$^4$ and Hermite-Gaussian beam emission. These results establish superlattice perturbations as a strategy for topological beam shaping and offer a route toward high-Q topological surface emitters.
\end{abstract}

%Use showkeys class option if keyword
                              %display desired

\maketitle
\raggedbottom % Prevent REVTeX from stretching paragraph gaps to equalize column heights.
%\cite{shen2017topological}.

Topological photonics has established robust guided interface states~\cite{khanikaev2017TwodimensionalTopologicalPhotonics,tang2022TopologicalPhotonicCrystals,2017NatureMaterdongValleyPhotonicCrystals} in systems where radiative leakage is negligible and a Hermitian description is applicable~\cite{luTopologicalPhotonics2014,ozawaTopologicalPhotonics2019,2009NaturewangObservationUnidirectionalBackscatteringimmune,wuSchemeAchievingTopological2015,shalaev2019RobustTransport}.
Many photonic platforms, including gratings, metasurfaces, and finite-thickness photonic-crystal slabs, are, however, open systems: their modes above the light line can couple to the free-space continuum and acquire complex eigenfrequencies~\cite{wang1993GuidedModeResonance,2002Phys.Rev.BtikhodeevQuasiguidedModesOptical,fan2002AnalysisGuidedResonances,2006Phys.Rev.BandreaniPhotoniccrystalSlabsTriangular,huang2023ResonantLeakyModes}.
Accounting for this radiative coupling requires a non-Hermitian description of their band structure and topology~\cite{shenTopologicalBandTheory2018,gongTopologicalPhasesNonHermitian2018,kawabataSymmetryTopologyNonHermitian2019,bergholtzExceptionalTopologyNonHermitian2021,poshakinskiyRadiativeTopologicalStates2014,2015NaturezhenSpawningRingsExceptional,nguyen2025GeneralizedNonHermitianHamiltonian,2026Phys.Rev.Lett.moritakeObservationRadiationLossBasedNonHermitian}.

Surface-emitting (SE) topological interface states in such open systems combine localization at domain walls with coupling to free-space radiation~\cite{2018NatCommungorlachFarfieldProbingLeaky,2020Sci.Adv.parappurathDirectObservationTopological,2020Nat.Nanotechnol.gaoDiracvortexTopologicalCavities,2022Nat.Photon.yangTopologicalcavitySurfaceemittingLaser,li2026NonHermitianDiracVortex}.
One-dimensional Jackiw-Rebbi states (JRs)~\cite{jackiw1976SolitonsFermion,2016asbothShortCourseTopological} in metagratings provide a minimal setting for studying these resonances~\cite{lee2021TopologicalGuidedMode,lee2021Erratum,choiTopologicalGuidedmodeResonances2025}, with implementations in dielectric gratings~\cite{2022Sci.Adv.leeTopologicalBeamingLight,2025LightSciApplchoiTopologicalBeamingLight}, metallic gratings~\cite{fu2026HighlyLocalizedPlasmonic}, and gratings based on two-dimensional materials~\cite{randerson2026TopologicalJackiwRebbiStates}.
A central challenge is to control both their radiative Q factors and far-field emission patterns, which are essential for lasing, polaritonic, and other applications.
The radiative lifetime is encoded in the imaginary part of the complex eigenfrequency~\cite{fan2003TemporalCoupledmodeTheory,wonjoosuh2004TemporalCoupledmodeTheory,2009J.Phys.A:Math.Theor.rotterNonHermitianHamiltonOperator,bykov2019CoupledwaveFormalismBound,2022Sci.Adv.leeTopologicalBeamingLight,randerson2026TopologicalJackiwRebbiStates}. Existing SE-JR implementations typically form topologically distinct domains by varying the filling fraction~\cite{leeBandFlipsBoundstate2019,lee2021TopologicalGuidedMode,2022Sci.Adv.leeTopologicalBeamingLight,2025LightSciApplchoiTopologicalBeamingLight,randerson2026TopologicalJackiwRebbiStates}, which generally modifies both Bragg coupling and out-of-plane radiation, linking changes in topological order to spectral detuning and radiative mismatch~\cite{2025LightSciApplchoiTopologicalBeamingLight,lee2021MetasurfacesBoundStates}.
Although flat-top beam shaping and numerical optimization of the Q factor has been discussed~\cite{2022Sci.Adv.leeTopologicalBeamingLight,2025LightSciApplchoiTopologicalBeamingLight,vakulenko2023AdiabaticTopologicalPhotonic}, quantitative control of both emission and radiative loss remains challenging. There is no unified non-Hermitian framework that explicitly relates the domain-wall profile to both the $Q$ factor and the emitted beam shape of SE states and provides practical guidance for engineering non-Hermitian topological photonic structures with SE states.

In this Letter, we develop a non-Hermitian Dirac formalism for SE-JRs in dielectric metagratings with second-order Brillouin-zone folding.
We show that the local superlattice perturbation spatial distribution shapes the complex Dirac mass profile, enabling control over spatial localization, radiative Q factors, and far-field emission while preserving the topological character of the interface states.
The theory provides analytical expressions for the emitted-field profile and the $Q$ factor, including their dependence on the perturbation amplitude and domain-wall geometry.
Applying this framework to a spatially graded superlattice perturbation, we obtain interface states with Gaussian near-field envelopes, first-order Hermite-Gaussian emission, and Q factors exceeding $10^4$.

\begin{figure}[tbp]
    \centering
    \includegraphics[width=\columnwidth]{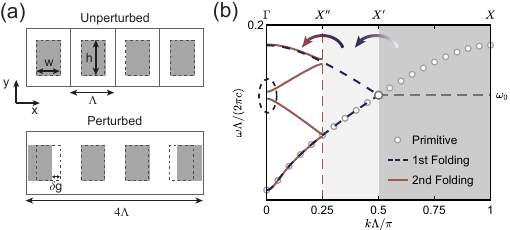}
    \caption{
    \textbf{Second-order Brillouin-zone folding in the metagrating.}
    \textbf{(a)} Schematics of the unperturbed grating (top) and gap-perturbed superlattice (bottom).
    Four primitive cells with period $\Lambda$, ridge width $w$, and thickness $h$ form a supercell of period $\Lambda_{\mathrm{s}}=4\Lambda$.
    The gap perturbation $\delta g$ displaces the two outer ridges in opposite directions while leaving the two central ridges fixed.
    Dashed outlines indicate the unperturbed ridge positions.
    \textbf{(b)}
    Calculated TE band structures for the unperturbed primitive grating and the gap-perturbed $4\Lambda$ supercell.
    Gray circles denote the fundamental $\mathrm{TE}_1$ band of the primitive grating.
    For clarity, the second-order folding is represented as two successive first-order folds.
    }
    \label{fig:1}
\end{figure}

We start our consideration from the second-order Brillouin-zone folding metagrating shown in Fig.~\ref{fig:1}(a). The unperturbed silicon grating is embedded in air and has a thickness of $h=340\,\mathrm{nm}$, a primitive period of $\Lambda=320\,\mathrm{nm}$, and a ridge width of $w=200\,\mathrm{nm}$. Four primitive cells are grouped into a supercell with period $4\Lambda$. The perturbation is introduced by keeping the two central ridges fixed while applying mirror-related displacements, parameterized by $\delta g$, to the two outer ridges.
A nonzero $\delta g$ breaks the primitive translational symmetry while preserving the central mirror symmetry. For isolated bands, the preserved mirror symmetry quantizes the Zak phase, enabling a $\mathbb{Z}_2$ classification of the folded bands~\cite{2022Phys.Rev.BtsubotaSymmetryprotectedQuantizationComplex}.

Figure~\ref{fig:1}(b) illustrates the corresponding second-order Brillouin-zone folding (BZF).
The fundamental transverse-electric (TE) band, characterized by the out-of-plane electric-field component $E_z$, of the primitive grating (gray circles) is folded into the reduced Brillouin zone (red line).
The band structures are calculated using the COMSOL eigenfrequency solver. For clarity, we represent the direct folding from period $\Lambda$ to period $4\Lambda$ as two successive first-order zone foldings, each corresponding to a doubling of the unit-cell period.
In the first folding step, a $2\Lambda$ unit cell containing two primitive cells is introduced, folding the primitive band into the reduced $\Gamma$-$X'$ region.
The counterpropagating Bloch states $\ket{+}$ and $\ket{-}$ at $k_{\pm}=\pm\pi/(2\Lambda)$ are degenerate at the reduced-zone boundary $X'$ with  eigenfrequency $\omega_0$, forming a Dirac point.
In the second folding step, the period is doubled to $4\Lambda$, yielding a supercell containing four primitive cells.
The Dirac point formed by $\ket{\pm}$ is folded from $X'$ to the zone center $\Gamma$.
A nonzero perturbation $\delta g$ then opens a Dirac gap at $\Gamma$, as highlighted by the dashed ellipse. The modulation enables coupling between the guided states $\ket{\pm}$ and the free-space radiation continuum, leading to complex eigenfrequencies and a non-Hermitian description of the folded-band dynamics.

\begin{figure}[tbp]
    \centering
    \includegraphics[width=\columnwidth]{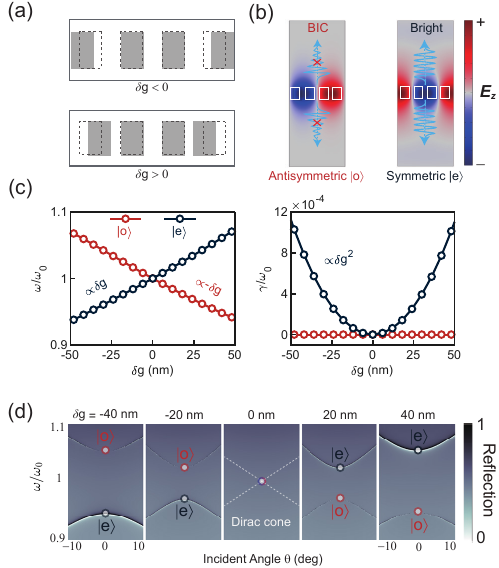}
\caption{\textbf{BZF-induced band inversion.}
\textbf{(a)} Sign convention for the mirror-related gap displacement $\delta g$.
\textbf{(b)} Simulated $E_z$ distributions of the two folded modes at the $\Gamma$ point for $\delta g=40\,\mathrm{nm}$.
The antisymmetric mode $\ket{o}$ forms an SP-BIC.
The symmetric mode $\ket{e}$ is bright.
\textbf{(c)} Real and imaginary parts of the two complex eigenfrequencies as functions of $\delta g$.
The antisymmetric mode $\ket{o}$ remains non-radiative and $\ket{e}$ bright.
Reversing $\delta g$ exchanges the mode ordering, while their center frequency remains $\omega_0$.
Accordingly, $m_{\mathrm{R}}=(\omega_{\mathrm{e}}-\omega_{\mathrm{o}})/2\propto\delta g$ and $m_{\mathrm{I}}=(\gamma_{\mathrm{e}}-\gamma_{\mathrm{o}})/2\propto\delta g^2$.
\textbf{(d)} Angle-resolved reflectance spectra calculated using RCWA for $\delta g=-40$, $-20$, $0$, $20$, and $40\,\mathrm{nm}$.}
\label{fig:2}
\end{figure}

\begin{figure*}[!t]
    \centering
    \includegraphics[width=\linewidth]{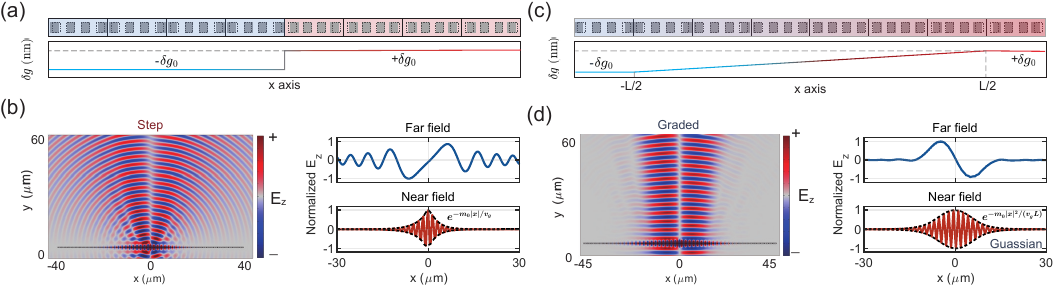}
\caption{\textbf{Spatial property of BZF-induced SE-JRs.}
\textbf{(a)} Schematic of a step domain wall formed by joining two superlattices with opposite perturbations, $-\delta g_0$ and $+\delta g_0$.
\textbf{(b)} Simulated electric-field distribution $E_z$ (left) and normalized far-field (upper right) and near-field (lower right) profiles.
The near-field profile is inside the slab and exhibits an exponentially decaying envelope away from the interface.
The emitted-field profile at $y = 50\,\mu\mathrm{m}$ featuring a central radiation node is observated.
\textbf{(c)} Schematic of a graded domain wall, where the perturbation $\delta g(x)$ varies linearly across the transition region of total length $L$, spanning $-L/2\leq x\leq L/2$ and connecting two superlattices with opposite perturbations, $-\delta g_0$ and $+\delta g_0$.
\textbf{(d)} Simulated electric-field distribution $E_z$ (left) and normalized far-field (upper right) and near-field (lower right) profiles.
The near-field profile is sampled inside the slab and exhibits a Gaussian envelope.
The far-field profile, evaluated $50\,\mu\mathrm{m}$ above the slab, exhibits an $\mathrm{HG}_1$ profile.}
\label{fig:3}
\end{figure*} 
%%%%%%%%%%%%%%%%%%%%%%%%%%%%%%%%%%%%%%%%%%%%%%%%%%%%%%%%%%%%%%%%%%%%%%%%%%%%%%%%%%%%%%%%%%%%%%%%%%%%%%%%%%%%%%%%%%%%%%%%%%%%%%%%%%%%%%%%%%%
To describe the real-frequency gap and the radiative dynamics of the folded bands, we construct an effective non-Hermitian Hamiltonian using perturbation theory: $H_{\mathrm{eff}}(q)=H_0(q)+\mathcal{V}$.
Here, $H_0(q)$ describes the counterpropagating guided modes of the primitive grating, while $\mathcal{V}$ accounts for the coherent coupling and radiative decay induced by the superlattice perturbation.
In the $\{\ket{+},\ket{-}\}$ basis, the non-Hermitian Hamiltonian takes the form:
\begin{equation}
    \begin{aligned}
        H_0(q)
        &=
        \omega_0\sigma_0+v_{\mathrm{g}}q\sigma_z,\\
        \mathcal{V}
        &=
        \begin{pmatrix}
            -\iu\gamma_+
            &
            \kappa-\iu\gamma_{\mathrm{g}}e^{\iu\varphi}
            \\
            \kappa-\iu\gamma_{\mathrm{g}}e^{-\iu\varphi}
            &
            -\iu\gamma_-
        \end{pmatrix}.
    \end{aligned}
    \label{eq:folded-perturbation-matrix}
\end{equation}
Here, $\omega_0$ is the frequency of the unperturbed folded degeneracy, $q$ is the reduced Bloch wavevector measured from the supercell $\Gamma$ point, and $v_{\mathrm{g}}$ is the unperturbed group velocity.
$\sigma_0=I_2$ is the identity matrix, and $\sigma_x$, $\sigma_y$, and $\sigma_z$ are the Pauli matrices.
The parameter $\kappa$ denotes the coherent coupling coefficient.
The quantities $\gamma_{\pm}$ are the diagonal radiative decay rates, while $\gamma_{\mathrm{g}}$ and $\varphi$ specify the magnitude and phase of the off-diagonal radiative coupling, respectively~\cite{koshelev2021BoundStatesContinuum}.

The displacement $\delta g$ can be modeled by a periodic effective permittivity perturbation $\Delta\varepsilon(x)$, which can be expanded in spatial harmonics as $\Delta\varepsilon(x)=\sum_n\Delta\varepsilon_n e^{\iu nGx}$, where  $G=2\pi/(4\Lambda)$ is the reduced reciprocal lattice vector and $\Delta\varepsilon_n$ denotes the Fourier coefficient of the $n$th spatial harmonic with wavevector $nG$.
The coherent coupling between the two folded guided modes is governed by the second spatial harmonic ($n=2$), with $\kappa\propto\bra{-}\Delta\varepsilon\ket{+}\propto\Delta\varepsilon_2$.
The radiative coupling instead involves the first-order spatial harmonics ($n=\pm1$), with $\gamma_+=\gamma_-=\gamma_{\mathrm{g}}\equiv\gamma_0=\frac{1}{2}\abs{d_0}^2\propto\abs{\Delta\varepsilon_{\pm1}}^2$ and $\varphi=0$.
Here, $d_0\propto\bra{0}\Delta\varepsilon\ket{\pm}$ represents the coupling amplitude between the folded guided modes and the zeroth-order radiation channel~\cite{2025LightSciApplchoiTopologicalBeamingLight}.

The effective Hamiltonian therefore reduces to
\begin{equation}
    \begin{aligned}
        H_{\mathrm{eff}}(q)
        =
        \Omega_0\sigma_0
        +
        v_{\mathrm{g}}q\sigma_z
        +
        m\sigma_x,
    \end{aligned}
    \label{eq:symmetry-reduced-folded-tcmt}
\end{equation}
in the form of the non-Hermitian Dirac Hamiltonian~\cite{shen2017topological}.
Here, $\Omega_0=\omega_0-\iu\gamma_0$ and $m=m_{\mathrm{R}}-\iu m_{\mathrm{I}}=\kappa-\iu\gamma_0$ denote the complex central eigenfrequency and the complex Dirac mass, respectively.

At the $\Gamma$ point, the eigenstates of Eq.~\eqref{eq:symmetry-reduced-folded-tcmt} split into symmetric $\ket{e}=(\ket{+}+\ket{-})/{\sqrt{2}}$ and antisymmetric $\ket{o}=(\ket{+}-\ket{-})/{\sqrt{2}}$, as shown in Fig.~\ref{fig:2}(b).
The antisymmetric mode $\ket{o}$ is a symmetry-protected bound state in the continuum (SP-BIC)~\cite{hsu2016BoundStatesContinuum,gao2016FormationMechanism,sadrieva2017TransitionOpticalBound}, as the odd field parity forbids coupling to the mirror-even zeroth-order radiation channel.
Its eigenfrequency is purely real: $\Omega_{\mathrm{o}}=\omega_0-m_{\mathrm{R}}$.
By contrast, the symmetric mode $\ket{e}$ is radiatively bright, with $\Omega_{\mathrm{e}}=\omega_0+m_{\mathrm{R}}-2\iu\gamma_0$.

In the weak-perturbation regime, the effective permittivity perturbation \(\Delta \varepsilon\) is linear in the geometric displacement $\delta g$.
Consequently, the real and imaginary parts of the Dirac mass exhibit linear and quadratic perturbation dependences ($m_{\mathrm{R}}\propto\delta g$ and $m_{\mathrm{I}}\propto\abs{\delta g}^2$), respectively.
Reversing $\delta g$ to \(-\delta g\) reverses $m_{\mathrm{R}}$ while preserving $m_{\mathrm{I}}$ and the central eigenfrequency $\Omega_0$ [the sign convention for $\delta g$ is illustrated in Fig.~\ref{fig:2}(a)]. 

Figure~\ref{fig:2}(c) shows the variation of the complex eigenfrequencies of these modes with $\delta g$.
Linear and quadratic dependences of the real and imaginary band gaps, $\omega_{\mathrm{e}}-\omega_{\mathrm{o}}=2m_{\mathrm{R}}\propto\delta g$
and
$\gamma_{\mathrm{e}}-\gamma_{\mathrm{o}}=2m_{\mathrm{I}}\propto\abs{\delta g}^2$, respectively, are observed, consistent with the above perturbation theory.
The reflectance spectra calculated using rigorous coupled-wave analysis (RCWA) provide the optical characterization of the band inversion, as shown in Fig.~\ref{fig:2}(d).
The reversal of the bright and dark band edges corroborates the sign reversal of the real component of the Dirac mass.

%%%%%%%%%%%%%%%%%%%%%%%%%%%%%%%%%%%%%%%%%%%%%%%%%%%%%%%%%%%%%%%%%%%%%%%%%%%%%%%%%%%%%%%%%%%%%%%%%%%%%
The interface between domains with different perturbations defines a domain wall.
It is characterized by a spatially varying complex Dirac mass $m(x)$ and a complex central eigenfrequency $\Omega_0(x)=\omega_0-\iu\gamma_0(x)$.
Substituting $q\rightarrow-\iu\partial_x$ into Eq.~\eqref{eq:symmetry-reduced-folded-tcmt} gives the real-space Dirac equation governing the interface state,
\begin{equation}
\left[
    -\iu v_{\mathrm{g}}\sigma_z\partial_x
    + m(x)\sigma_x
\right]\Psi(x)
=
\left(\Omega-\Omega_0(x)\right)\Psi(x).
\label{eq:jr-real-space-dirac}
\end{equation}
Under the first order weak radiative loss approximation, the corresponding eigenmode is given by
\begin{equation}
 \Psi(x)\simeq\psi_0(x)
 \left\{[1+a(x)]\chi_-+b(x)\chi_+\right\},
 \label{eq:perturbative-state}
\end{equation}
with a complex eigenfrequency \(\Omega_{\mathrm{JR}}\simeq\omega_0-\iu\Gamma_{\mathrm{JR}}\), where \(\Gamma_{\mathrm{JR}}=\int_{-\infty}^{+\infty}\gamma_0(x)\psi_0^2(x)\,\mathrm{d}x\).
This eigenmode is normalizable only when the real Dirac masses in the two domains have opposite signs, $m_{\mathrm{R}}(-\infty)m_{\mathrm{R}}(+\infty)<0$.
Here, the normalized spatial envelope of the Hermitian Dirac solution is \(\psi_0(x)=\mathcal{N}_0\exp\!\left[-\int_0^xm_{\mathrm{R}}(x')/v_{\mathrm{g}}\,\mathrm{d}x'\right]\), where $\mathcal{N}_0$ is a normalization constant. 
The normalized spinors \(\chi_-=(1,-\iu)^{\mathsf{T}}/\sqrt{2}\) and \(\chi_+=(1,\iu)^{\mathsf{T}}/\sqrt{2}\) form a basis of opposite chirality.
The non-Hermitian corrections are given by
\begin{equation}
\begin{aligned}
a(x)&=\iu\int_0^x\frac{m_{\mathrm{I}}(x')}{v_{\mathrm{g}}}\,\mathrm{d}x',\\
b(x)&=\frac{1}{v_{\mathrm{g}}\psi_0^2(x)}\int_{-\infty}^x
[\Gamma_{\mathrm{JR}}-\gamma_0(x)]\psi_0^2(s)\,\mathrm{d}x.
\end{aligned}
\end{equation}
The coefficient \(a(x)\) describes the phase accumulation induced by the imaginary mass \(m_{\mathrm{I}}(x)\). And the coefficient \(b(x)\) gives the opposite-chirality admixture induced by spatially varying radiative leakage \(\gamma_0(x)\).

The electric field envelope inside the grating \({E}(x)\) and radiative aperture \(A(x)\) are reconstructed by superposing the counterpropagating guided-mode profiles,
\begin{equation}
    \begin{aligned}
        E(x) &= \mathcal{E}_+\Psi_+(x)+\mathcal{E}_-\Psi_-(x),\\
        A(x) &= \mathcal{E}_+d_+(x)\Psi_+(x)+\mathcal{E}_-d_-(x)\Psi_-(x).
    \end{aligned}
\label{eq:jr-near-field}
\end{equation}
Here $\Psi_\pm(x)$ are the components of $\Psi(x)$ in the $\{\ket{+},\ket{-}\}$ basis; ${\mathcal{E}}_\pm$ and $d_\pm(x)$ denote the guided electric-field envelopes and the radiation coupling coefficients of $\ket{+}$ and $\ket{-}$, respectively.
As discussed above, mirror symmetry implies \({\mathcal{E}}_+={\mathcal{E}}_-=\mathcal{E}_0\) and \(d_+(x)=d_-(x)=d_0(x)\).
A constant overall phase is absorbed into the field amplitude in the simplified profiles below. 
Details of the derivation of the near field and radiative aperture are provided in the Supplemental Material~\cite{SuppMat}.

We first consider a step interface consisting of superlattice gratings with opposite perturbations, $-\delta g_0$ and $+\delta g_0$, as illustrated in Fig.~\ref{fig:3}(a).
The real and imaginary components of the Dirac mass satisfy $m_{\mathrm{R}}(x)=m_0\operatorname{sgn}(x)$ and $m_{\mathrm{I}}(x)=\gamma_0(x)=\gamma_0$.
Here $m_0>0$ denotes the magnitude of the real Dirac mass in the outer domains and $\gamma_0$  is the corresponding radiative decay rate.
The coefficient \(b(x)\) in Eq.~\eqref{eq:perturbative-state} vanishes and the interface mode is given by \(\Psi_{\mathrm{step}}(x)=\mathcal{N}_{\mathrm{step}} e^{-m_0\abs{x}/v_{\mathrm{g}}} e^{\iu \gamma_0x/v_{\mathrm{g}}}\chi_-\).
The radiation coupling coefficient $d_0(x)=d_0\operatorname{sgn}(x)$, where $d_0$ denotes the radiative coupling coefficient at $\delta g=+\delta g_0$. 

The corresponding near-field envelope and radiative aperture can be expressed as:
\begin{equation}
    \begin{aligned}
         E_{\mathrm{step}}(x) &= \mathcal{N}_{\mathrm{step}}\mathcal{E}_0 e^{-m_0\abs{x}/v_{\mathrm{g}}}e^{\iu \gamma_0x/v_{\mathrm{g}}},\\
    A_{\mathrm{step}}(x)&= \mathcal{N}_{\mathrm{step}}\mathcal{E}_0d_0\operatorname{sgn}(x)e^{-m_0\abs{x}/v_{\mathrm{g}}} e^{\iu \gamma_0x/v_{\mathrm{g}}}.
    \end{aligned}
\end{equation}
The near-field envelope decays exponentially away from the interface, consistent with the simulated field distribution in Fig.~\ref{fig:3}(b).
The sign reversal of $\delta g$ imposes a $\pi$ phase jump in the radiation coupling across the interface.
This phase jump suppresses emission in the surface-normal direction in the weak radiative loss approximation (details in Supplemental Material~\cite{SuppMat}), as shown in the upper right panel of Fig.~\ref{fig:3}(b).

\begin{figure}[tbp]
    \centering
    \includegraphics[width=\columnwidth]{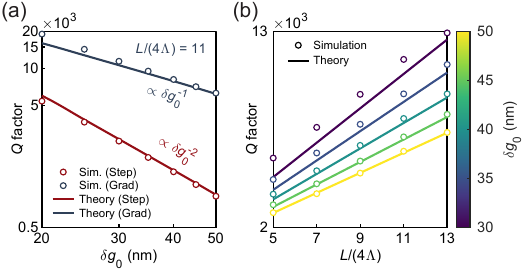}
    \caption{\textbf{Temporal confinement of BZF-induced SE-JRs.}
\textbf{(a)} $Q$ factors as a function of the perturbation amplitude $\delta g_0$ for step (red) and graded (blue) interfaces. The graded junction has a fixed length of $L=11\Lambda_{\mathrm{s}}$.
The step and graded interfaces exhibit $Q\propto(\delta g_0)^{-2}$ and $Q\propto(\delta g_0)^{-1}$, respectively.
\textbf{(b)} $Q$ factor of the graded SE-JR as a function of the normalized junction length $L/(4\Lambda)$ at $\delta g_0=30\text{ to }50\,\mathrm{nm}$, consistent with the predicted linear scaling $Q_{\mathrm{grad}}\propto L$.
Solid lines show theoretical predictions evaluated using bulk parameters extracted from calculations with periodic boundary conditions, and circles denote eigenfrequency simulation results.
}
\label{fig:4}
\end{figure}

The eigenfrequency of the SE-JRs at the step interface lies at the center of the complex gap, $\Omega_{\mathrm{step}}=\Omega_0$, giving the $Q$ factor
\begin{equation}
    Q_{\mathrm{step}}=\frac{\omega_0}{2\gamma_0}\propto(\delta g_0)^{-2}.
\end{equation}
Figure~\ref{fig:4}(a) compares eigenfrequency simulations of the step heterostructure (red circles) with the theoretical prediction (red line). 
The numerical method is described in the Supplemental Material~\cite{SuppMat}.
%%%%%%%%%%%%%%%%%%%%%%%%%%%%%%%%%%%%%%%%%%%%%%%%%%%%%%%%%%%%%%%%%%%%%%%%%%%%%%%%%%%%%%%%%%%%%%%%%%%%%%%%%%%%%%%%%%%%%%%%%%%%%%%%%%%%%%%%%%%%%%%%%%%%%%%%%%%%%%%%%%%%%%%%%%%%%%%%%%%%%%%

The continuously graded domain wall shown in Fig.~\ref{fig:3}(c) provides additional control over both spatial and temporal properties.
Here, $L$ denotes the total junction length, spanning $-L/2\leq x\leq L/2$.
The perturbation varies linearly as $\delta g(x)=2\delta g_0 x/L$ within the region $\abs{x}\leq L/2$ and saturates in the two outer domains at $\delta g(x)=\delta g_0\operatorname{sgn}(x)$ in the domain $\abs{x}>L/2$.

Within the graded region, this linear perturbation gives $m_{\mathrm{R}}(x)=2m_0x/L$, $m_{\mathrm{I}}(x)=4\gamma_0x^2/L^2$ and \(d_0(x)=2d_0x/L\).
The corresponding Hermitian envelope is expressed as $\psi_{0,\mathrm{grad}}(x)=\mathcal{N}_{\mathrm{grad}} e^{-x^2/(2\xi^2)}$, where $\xi=\sqrt{v_{\mathrm{g}}L/(2m_0)}$ is the characteristic localization length of the interface state. 

The corresponding near-field and radiative aperture are given by
\begin{equation}
\begin{aligned}
  E_{\mathrm{grad}}(x) &\simeq\mathcal{N}_{\mathrm{grad}}\mathcal{E}_0 e^{-x^2/(2\xi^2)},\\
  A_{\mathrm{grad}}(x) &\simeq \frac{2\mathcal{N}_{\mathrm{grad}}\mathcal{E}_0d_0}{L}x e^{-x^2/(2\xi^2)}.
\end{aligned}
\label{eq:jr-continuous-near-field}
\end{equation}
The near-field envelope exhibits a Gaussian profile, whereas the radiative aperture approaches an $\mathrm{HG}_1$ profile as shown in Fig.~\ref{fig:3}(d).

To evaluate the eigenfrequency, we futher assume the mode is predominantly confined to the graded region, with a localization length $\xi\ll L/2$.
The corresponding complex eigenfrequency is given by:
\(\Omega_{\mathrm{grad}}\simeq\omega_0-\iu{2\gamma_0\xi^2}/{L^2}\) (see the Supplemental Material~\cite{SuppMat} for details).
The corresponding $Q$ factor is
\begin{equation}
    Q_{\mathrm{grad}}\simeq\frac{\omega_0 m_0 L}{2\gamma_0 v_{\mathrm{g}}}\propto L(\delta g_0)^{-1}.
\end{equation}
Figure~\ref{fig:4}(a) compares the analytically predicted $Q$ factors with those obtained from simulations of the graded heterostructure with $L/(4\Lambda)=11$, illustrating the inverse dependence of the $Q$ factor on the perturbation amplitude \(\delta g_0\).
Figure~\ref{fig:4}(b) presents the corresponding comparison as a function of the normalized grading length $L/(4\Lambda)$ for perturbation amplitudes between $30$ and $50\,\mathrm{nm}$.
At fixed $\delta g_0$, the simulations are consistent with the predicted linear increase of $Q_{\mathrm{grad}}$ with $L$.
This increase reflects a reduction in the mode-averaged radiative loss as the grading length increases.
The simulated graded structures attain $Q$ factors exceeding $10^4$, demonstrating that spatial grading enables high-$Q$ surface-emitting interface states.

In summary, we have developed a non-Hermitian Dirac framework for SE-JRs in dielectric metagratings with second-order Brillouin-zone folding. By shaping the complex Dirac mass through the local amplitude of a superlattice perturbation, we establish a quantitative connection between the domain-wall profile, spatial localization, radiative quality factor, and emitted beam shape while preserving the topological character of the interface states. A linearly graded perturbation supports states with quality factors exceeding $10^4$ and first-order Hermite-Gaussian emission. The theory predicts a linear increase in the $Q$ factor with the graded junction length, providing an explicit design rule for enhancing temporal confinement. Achieving higher $Q$ factors through longer junctions and weaker perturbations nevertheless requires precise geometric control, introducing a practical trade-off between radiative confinement and fabrication tolerance. These results establish superlattice perturbations as a strategy for topological beam shaping and offer a route toward high-$Q$ topological surface emitters, with potential extensions to two-dimensional structures supporting more complex emission profiles.

\textit{Acknowledgment\,--\,}The authors thank Maxim Gorlach, Meng Xiao, Bailing Zhang and Alexander Poddubny for valuable discussions and insightful suggestions. This work was supported by the Russian Science Foundation (Grant No. 25-79-31027), the National Natural Science Foundation of China (Grant No. W2532010) and the Academic Leadership Program Priority 2030.

\textit{Data availability\,--\,}The data that support the findings of this Letter are not publicly available. The data are available from the authors upon reasonable request.

\bibliography{ref}
\end{document}